\documentclass[11pt]{article}
\usepackage{amsmath}
\usepackage{amsfonts}
\usepackage{amsthm}
\usepackage{fullpage}
\usepackage[boxruled,scleft,vlined]{algorithm2e}
\usepackage{enumitem}

\newcommand{\old}[1]{#1_{\mathit{old}}}

\newcommand{\reals}{\ensuremath{\mathbb{R}}}

\newtheorem{theorem}{Theorem}[section]
\newtheorem{lemma}[theorem]{Lemma}
\newtheorem{corollary}[theorem]{Corollary}
\theoremstyle{definition}
\newtheorem{definition}{Definition}[section]

\newcommand{\ground}{\mathcal{G}}
\newcommand{\N}{N}
\newcommand{\I}{\mathcal{I}}
\newcommand{\sinit}{S_\mathit{init}}

\SetKw{Break}{break}

\author{Justin Ward \\
\tt jward@cs.toronto.edu \\
\newline \small Dept. of Computer Science, University of Toronto, Toronto, Canada
}

\title{A $(k + 3)/2$-approximation algorithm for monotone submodular $k$-set packing and general $k$-exchange systems}
\date{\today}

\begin{document}
\maketitle

\begin{abstract}
We consider the monotone submodular $k$-set packing problem in the context of the more general problem of maximizing a monotone submodular function in a $k$-exchange system.  These systems, introduced by Feldman et al.\ \cite{Feldman-2011}, generalize the matroid k-parity problem in a wide class of matroids and capture many other combinatorial optimization problems.  We give a deterministic, non-oblivious local search algorithm that attains an approximation ratio of $(k + 3)/2 + \epsilon$ for the problem of maximizing a monotone submodular function in a $k$-exchange system, improving on the best known result of $k + \epsilon$, and answering an open question posed in Feldman et al.
\end{abstract}

\section{Introduction}

In the general $k$-set packing problem, we are given a collection $\ground$ of sets, each with at most $k$ elements, and an objective function $f : 2^\ground \to \reals_+$ assigning each subset of $\ground$ a value, and seek a collection of pairwise-disjoint sets $S \subseteq \ground$ that maximizes $f$.  In the special case that $f(A) = |A|$, we obtain the \emph{unweighted $k$-set packing problem}.  Similarly, if $f$ is linear function, so that $f(A) = \sum_{e \in A}w(e)$ for some weight function $w : \ground \to \reals_+$ we obtain the \emph{weighted $k$-set packing problem}.  In this paper we consider the case in which $f$ may be any monotone submodular function.

For unweighted $k$-set packing, Hurkens and Schrijver \cite{Hurkens-1989} and Halld\'orsson \cite{Halldorsson-1995} independently obtained a $k/2 + \epsilon$ approximation via a simple local search algorithm.
Using similar techniques, Arkin and Hassin \cite{Arkin-1997} obtained a $k - 1 + \epsilon$ approximation for weighted $k$-set packing, and showed that this result is tight for their simple local search algorithm.  Chandra and Halld\'orsson \cite{Chandra-1999} showed that a more sophisticated local search algorithm, which starts with a greedy solution and always chooses the best possible local improvement at each stage, attains an approximation ratio of $2(k + 1)/3 + \epsilon$.  This was improved further by Berman \cite{Berman-2000}, who gave a \emph{non-oblivious} local search algorithm yielding a $(k + 1)/2 + \epsilon$ approximation for weighted $k$-set packing.  Non-oblivious local search \cite{Khanna-1994} is a variant of local search in which an auxiliary objective function to evaluate solutions, rather than the problem's given objective.  In the case of Berman, the local search procedure repeatedly seeks to improve the sum of \emph{square} of the weights in the current solution, rather than the sum of the weights.

Many of the above local search algorithms for $k$-set packing yield the same approximations for the more general problem of finding maximum independent sets in $(k + 1)$-claw free graphs.  Additionally, local search techniques have proved promising for other generalizations of $k$-set packing, including variants of the matroid $k$-parity problem \cite{Lee-2010,Soto-2011a}.  Motivated by the similarities between these problems, Feldman et al.\ \cite{Feldman-2011} introduced the class of \emph{$k$-exchange} systems, which captures problems amenable to approximation by local search algorithms.  These systems are formulated in the general language of independence systems, which we now briefly review.

An independence system is specified by a ground set $\ground$, and a hereditary (i.e.\ non-empty and downward-closed) family $\I$ of subsets of $\ground$.  These subsets of $\ground$ contained in $\I$ are called \emph{independent sets}, and the inclusion-wise maximal sets of $\I$ are called \emph{bases} of the independence system $(\I, \ground)$.  Given an independence system $(\ground, \I)$ and a function $f : 2^\ground \to \reals_+$, we consider the  problem of finding an independent set $S \in \I$ that maximizes $f$.  

The class of $k$-exchange systems satisfy the following additional property:
\begin{definition}[$k$-exchange system \cite{Feldman-2011}]
\label{def:k-exchange}
A hereditary system $\I$ is a \emph{$k$-exchange system} if,
for all $A$ and $B$ in $\I$, there exists a multiset $Y = \{ Y_e \subseteq B \setminus A\ |\ e \in A \setminus B \}$, containing a subset $Y_e$ of $B \setminus A$ for each element $e \in A \setminus B$, that satisfies:
\begin{enumerate}[itemsep=0em,topsep=0.5em,leftmargin=0.65in,label={\rm (K\arabic{*})}]
\item $|Y_e| \le k$ for each $x \in A$. \label{N1}
\item Every $x \in B \setminus A$ appears in at most $k$ sets of $Y$. \label{N2}
\item For all $C \subseteq A \setminus B$, $(B \setminus \left(\bigcup_{e \in C}Y_e\right)) \cup C
  \in \I$. \label{N3}
\end{enumerate}
\end{definition}
We call the set $Y_e$ in Definition \ref{def:k-exchange} the \emph{neighborhood} of $e$ in $B$.  For convenience, we extend the collection $Y$ in Definition \ref{def:k-exchange} by including the set $Y_x = \{x\}$ for each element $x \in A \cap B$.  It is easy to verify that the resulting collection still satisfies conditions \ref{N1}--\ref{N3}.

The 1-exchange systems are precisely the class of strongly base orderable matroids described by Brualdi \cite{Brualdi-1971}.  This class is quite large and includes all gammoids, and hence all transversal and partition matroids.  For $k > 1$, the class of $k$-exchange systems may be viewed as a common generalization of the matroid $k$-parity problem in strongly base orderable matroids and the independent set problem in $(k + 1)$-claw free graphs.
Feldman et al. showed that $k$-exchange systems encompass a wide variety of combinatorial optimization problems, including as $k$-set packing, intersection of $k$ strongly base orderable matroids, hypergraph $b$-matching (here $k = 2$), as well as problems such as asymmetric traveling salesperson (here $k = 3$).

Our results hold for any $k$-exchange system, and so we present them in the general language of Definition \ref{def:k-exchange}.  However, the reader may find it helpful to think in terms of a concrete problem, such as the $k$-set packing problem.  In that case, the ground set $\ground$ is the given collection of sets, and a sub-collection of sets $S \subseteq \ground$ is independent if and only if all of the sets in $S$ are disjoint. Given $A$ and $B$ as in Definition \ref{def:k-exchange}, $Y_e$ is the set of all sets in $B$ that contain any element contained by the set $e \in A$ (i.e. the set of all sets in $B$ that are not disjoint from $e$).  Then, property \ref{N3} is immediate, and \ref{N1} and \ref{N2} follow directly from the fact that each set in $\ground$ contains at most $k$ elements.

\subsection{Related Work}

Recently, the problem of maximizing submodular functions subject to various constraints has attracted much attention.  We focus here primarily on results pertaining to matroid constraints and related independence systems.

In the case of an arbitrary single matroid constraint, Calinescu et al.\ have attained a $e/(e - 1)$ approximation for monotone submodular maximization, via the \emph{continuous greedy algorithm}.  This result is tight, provided that $P \neq NP$ \cite{Feige-1998}.  In the case of $k \ge 2$ simultaneous matroid constraints, an early result of Fisher, Nemhauser, and Wolsey \cite{Fisher-1978} shows that the standard greedy algorithm attains a $k + 1$ approximation for monotone submodular maximization.  Fischer et al.\ state further that the result can be generalized to $k$-systems (a full proof appears in Calinescu et al.\ \cite{Calinescu-2007}).  More recently, Lee, Sviridenko, and Vondr\`ak\ \cite{Lee-2010a} have improved this result to give a $k + \epsilon$ approximation for monotone submodular maximization over $k \ge 2$ arbitrary matroid constraints, via a simple, oblivious local search algorithm.
A similar analysis was used by Feldman et al.\ \cite{Feldman-2011} to show that oblivious local search attains a $k + \epsilon$ approximation for the class of $k$-exchange systems (here, again, $k \ge 2$).  For the more general class of $k$-systems, Gupta et al.\ \cite{Gupta-2010} give a $(1 + \beta)(k + 2 + 1/k)$ approximation, where $\beta$ is the best known approximation ratio for unconstrained non-monotone submodular maximization.

In the case of unconstrained non-monotone submodular maximization, 
Feige, Mirrokni, and Vondr\'ak \cite{Feige-2007} gave a randomized $2.5$ approximation, which was iteratively improved by Gharan and Vondr\'{a}k \cite{Gharan-2011} and then Feldman, Naor, and Shwartz \cite{Feldman-2011a} to $\approx 2.38$.  For non-monotone maximization subject to $k$ matroid constraints, Lee, Sviridenko, and Vondr\'ak \cite{Lee-2009} gave a $k + 2 + 1/k + \epsilon$ approximation, and later improved \cite{Lee-2010a} this to a $k + 1 + 1/(k - 1) + \epsilon$ approximation.  Again, the latter result is obtained by a standard local search algorithm.  Feldman et al.\ \cite{Feldman-2011} apply similar techniques to yield a $k + 1 + 1/(k - 1) + \epsilon$ approximation for non-monotone submodular maximization the general class of $k$-exchange systems.


\subsection{Our Contribution}

In the restricted case of a linear objective function, Feldman et al.\ \cite{Feldman-2011} gave a non-oblivious local search algorithm inspired by Berman's algorithm \cite{Berman-2000} for  $(k + 1)$-claw free graphs.  They showed that the resulting algorithm is a $(k + 1)/2 + \epsilon$ approximation for linear maximization in any $k$-exchange system.  Here we consider a question posed in \cite{Feldman-2011}: namely, whether a similar technique can be applied to the case of monotone submodular maximization in $k$-exchange systems.  In this paper, we give a successful application of the non-oblivious local search techniques to the case of monotone submodular maximization in a $k$-exchange system.  As in \cite{Feldman-2011}, the $k$-exchange property is used only in the analysis of our algorithm.  The resulting non-oblivious local search algorithm attains an approximation factor of $\frac{k + 3}{2} + \epsilon$.  For $k > 3$, this improves upon the $k + \epsilon$ approximation obtained by the oblivious local search algorithm presented in \cite{Feldman-2011}.  Additionally, we note that our algorithm runs in time polynomial in $\epsilon^{-1}$, while the $k + \epsilon$ approximation algorithm of \cite{Feldman-2011} requires time exponential in $\epsilon^{-1}$.

As a consequence of our general result, we obtain an improved approximation guarantee of $\frac{k + 3}{2}$ for a variety of monotone submodular maximization problems (some of which are generalizations of one another) including: $k$-set packing, independent sets in $(k + 1)$-claw free graphs, $k$-dimensional matching, intersection of $k$ strongly base orderable matroids, and matroid $k$-parity in a strongly base orderable matroid.  In all cases, the previously best known result was $k + \epsilon$.

\section{A First Attempt at the Submodular Case}

Before presenting our algorithm, we describe some of the difficulties that arise when attempting to adapt the non-oblivious local search algorithm of \cite{Feldman-2011} and \cite{Berman-2000} to the submodular case.  Our hope is that this will provide some intuition for our algorithm, which we present in the next section.

We recall that a function $f : 2^\ground \to \reals_+$ is submodular if $f(A) + f(B) \ge f(A \cup B) + f(A \cap B)$ for all $A, B \subseteq \ground$.  Equivalently, $f$ is submodular if for all $S \subseteq T$ and all $x \not\in T$, $f(S + x) - f(S) \ge f(T + x) - f(T)$.  In other words, submodular functions are characterized by decreasing marginal gains.  We say that a submodular function $f$ is monotone if it additionally satisfies $f(S) \le f(T)$ for all $S \subseteq T$.

The non-oblivious algorithm of \cite{Feldman-2011} for the linear case is shown in Algorithm \ref{alg:nols}.  It proceeds from one solution to another by applying a \emph{$k$-replacement} $(A,B)$.  Formally, we call the pair of sets $(A,B)$, where $A \subseteq \ground \setminus (S \setminus B)$ and $B \subseteq S$ a $k$-replacement if $|A| \le k$, $|B| \le k^2 - k + 1$ and $(S \setminus B) \cup A \in \I$.  Algorithm \ref{alg:nols} repeatedly searches for a $k$-replacement that improves an auxiliary potential function $w^2$.  Because $f$ is linear, it can be represented as a sum of weights, one for each element in $S$.   If $w(e) = f(\{e\})$ is the weight assigned to an element $e$ in this representation, then the non-oblivious potential function is given by $w^2(S) = \sum_{e \in S}w(e)^2$.  That is, our non oblivious potential function $w^2(S)$ is simply the sum of the \emph{squared} weights of the elements of  $S$.\footnote{To ensure polynomial-time convergence, Algorithm \ref{alg:nols} first round the weights down to integer multiples of a suitable small value $\alpha$, related to the approximation parameter $\epsilon$.  The algorithm then converges in time polynomial in $\epsilon^{-1}$ and $n$, at a loss of only $(1 - \epsilon)^{-1}$ in the approximation factor.}  
 \begin{algorithm}
\KwIn{\parbox[t]{7in}{\begin{itemize}[parsep=0em,itemsep=0em,topsep=0em,leftmargin=1em]
 \item Ground set $\ground$ 
 \item Membership oracle for $\mathcal{I} \subseteq 2^\ground$ 
 \item Value oracle for monotone submodular function $f : 2^\ground \to \reals_+$ 
 \item Approximation parameter $\epsilon \in (0,1)$
 \end{itemize}}}
 Let $\sinit = \{ \arg\max_{e \in \ground} w(e) \}$\;
 Let $\alpha = w(\sinit)\epsilon/n$\;
 Round all weights $w(e)$ down to integer multiples of $\alpha$\;
 $S \gets \sinit$\;
 $\old{S} \gets S$\;
 \Repeat{$\old{S} = S$}{
   \ForEach{$k$-replacement $(A,B)$}{
     \If{$w^2(A) > w^2(B)$}{
       $\old{S} \gets S$\;
       $S \gets (S \setminus B) \cup A$\;
       \Break\;
     }
   }
 }
 \Return $S$\;
\caption{Non-Oblivious Local Search for Linear Objective Functions}
\label{alg:nols}
\end{algorithm}

In the monotone submodular case, we can no longer necessarily represent $f$ as a sum of weights.  However, borrowing some intuition from the greedy algorithm, we might decide to replace each weight $w(e)$ in the potential function $w$ with the marginal gain/loss associated with $e$.  That is, at the start of each iteration of the local search algorithm, we assign each element $e \in \ground$ weight $w(e) = f(S + e) - f(S - e)$, where $S$ is the algorithm's current solution, then we proceed as before.  Note that $w(e)$ is simply the marginal gain attained by adding $e$ to $S$, in the case that $e \not\in S$ or the marginal loss suffered by removing $e$ from $S$, in the case that $e \in S$.  We define the non-oblivious potential function $w^2$ in terms of the (current) weight function $w$, as before.


Unfortunately, the resulting algorithm may fail to terminate, as the following small example shows.  We consider a simple, unweighted coverage function on the universe $U = \{a,b,c,x,y,z\}$, defined as follows. Let:
\begin{align*}
S_1 &= \{a, b \} & S_3 &= \{x, y \} \\
S_2 &= \{a, c \} & S_4 &= \{x, z \}
\end{align*}
We define $\ground = \{1,2,3,4\}$ and $f(A) = \left| \bigcup_{i \in A} S_i \right|$ for all $A \subseteq \ground$.  Finally, we consider the 2-exchange system with only 2 bases: $P = \{1,2\}$ and $Q = \{3,4\}$.  For current solution $S = P$ we have $w(1) = w(2) = 1$ and $w(3) = w(4) = 2$.  Since $w^2(\{1,2\}) = 2 < 8 = w^2(\{3,4\})$, the 2-replacement $(\{3,4\}, \{1,2\})$ is applied, and the current solution becomes $Q$.  In the next iteration, we have $S = Q$, and $w(1) = w(2) = 2$ and $w(3) = w(4) = 1$, so the 2-replacement $(\{1,2\}, \{3,4\})$ is applied by the algorithm.  This returns us to the solution to $P$, where the process repeats indefinitely.

\section{The New Algorithm}
Intuitively, the problem with this initial approach stems from the fact that the weight function used at each step of the algorithm depends on the current solution $S$ (since all marginals are taken with respect to $S$).  Each time the algorithm makes an improvement, it changes the current solution, thereby changing the weights assigned to all elements in the next iteration.  Hence, we are effectively making use of an entire family of non-oblivious potential functions, indexed by the current solution $S$.  It may be the case that a $k$-replacement $(A,B)$ results in an improvement with respect to the current solution's potential function, but in fact results in a \emph{decreased} potential value in the next iteration, after the weights have been updated.  

Surprisingly, we can solve the problem by introducing \emph{even more} variation in the potential function.  Specifically, we allow the algorithm to use a different weight function not only for each current solution $S$, but also for \emph{each $k$-replacement $(A,B)$ that is considered}.  We give the full algorithm at the end of this section and a detailed analysis in the next.



Our general approach is to consider the elements of a set $X$ in some order $\prec$, and assign to each $e \in X$ in the set the marginal gain in $f$ obtained when it is added to the set containing all the preceding elements of $X$.  By carefully updating $\prec$ together with the current solution $S$ at each step, we ensure that the algorithm converges to a local optimum.  We now give the details for how are weights are calculated, given the current ordering $\prec$.

At each iteration of the algorithm, before searching an improving $k$-replacement, we assign weights $w$ to all of the elements of the current solution $S$.  The weights will necessarily depend on $S$, but remain fixed for all $k$-replacements considered in the current phase.  Let $s_i$ be the $i$th element of $S$ in the ordering
$\prec$ and let $S_i = \{s_j \in S : j \le i\}$ be the set containing the first $i$ elements of $S$ in the ordering $\prec$.  Then, the weight function $w : S \to \reals_+$ is given by
\[
w(s_i) = f(S_{i - 1} + s_i) - f(S_{i - 1})
\]
for all $s_i \in S$.  Note that our weight function satisfies
\begin{equation}
  \label{eq:15}
  \sum_{x \in S}w(x) = \sum_{i = 1}^{|S|} f(S_{i - 1} + s_i) - f(S_{i - 1}) = f(S)
\end{equation}

In order to evaluate each $k$-replacement $(A,B)$, we need to assign weights to the elements in $A \subseteq \ground \setminus (S \setminus B)$.  We use a different weight function for each $k$-replacement.  Suppose that we are considering the $k$-replacement $(A,B)$.  Let $a_i$ be the $i$th element of $A$ in the ordering $\prec$ and let $A_i = \{a_j \in A : j \le i\}$ be the set containing the first $i$ elements of $A$ in the ordering $\prec$.  Then, we define the weight function $w_{(A,B)} : A \to \reals_+$ by
\[
w_{(A,B)}(a_i) = f((S \setminus B) \cup A_{i - 1} + a_i) - f((S \setminus B) \cup A_{i - 1})
\]
for all $a_i \in A$. Note that for every $k$-replacement $(A,B)$,
\begin{multline}
  \label{eq:16}p
  \sum_{x \in A}w_{(A,B)}(x) \ge \sum_{i = 1}^{|A|}\left(f((S \setminus B) \cup A_{i - 1} + a_i) - f((S \setminus B) \cup A_{i - 1})\right) \\ = f((S \setminus B) \cup A) - f(S \setminus B) \ge f(S \cup A) - f(S)
\enspace,
\end{multline}
where the last inequality follows from the decreasing marginals characterization of submodularity.  Note that since the function $f$ is \emph{monotone} submodular, all of the weights $w$ and $w_{(A,B)}$ that we consider will be nonnegative.  This fact plays a crucial role in our analysis.  

Our final algorithm appears in Algorithm \ref{alg:1}.  We start from an initial solution 
$\sinit = \{\arg\max_{e \in \ground} f(\{e\})\}$, consisting of the singleton set of largest value.  Note that after applying a $k$-replacement $(A,B)$, the algorithm updates $\prec$ to ensure that all of the elements of $S \setminus B$ precede those of $A$.  As we shall see in the next section, this ensures that the algorithm will converge to a local optimum.
As in the linear case, we use the non-oblivious potentials $w^2(B) = \sum_{b \in B}w(b)^2$ and $w_{(A,B)}(A) = \sum_{a \in A}w_{(A,B)}(a)^2$.  Again, we note that while \emph{all} of our weights implicitly depend on the current solution, the weights $w_{(A,B)}$ additionally depend on the $k$-replacement $(A,B)$ considered.

Additionally, to ensure polynomial-time convergence, we round all of our weights down to the nearest integer multiple of $\alpha$, depending on the parameter $\epsilon$.  This will ensure that every improvement improves the current solution by an additive factor of at least $\alpha^2$.
Because of this rounding factor, we must actually work with the following analogs of (\ref{eq:15}) and (\ref{eq:16}):
\begin{equation}
  \label{eq:15r}
  \sum_{x \in S}w(x) \le \sum_{i = 1}^{|S|} \left(f(S_{i - 1} + s_i) - f(S_{i - 1})\right) = \sum_{i = 1}^{|S|} \left(f(S_{i}) - f(S_{i - 1})\right) = f(S) - f(\emptyset) \le f(S)
\end{equation}
\begin{multline}
  \label{eq:16r}
  \sum_{x \in A}w_{(A,B)}(x) \ge \sum_{i = 1}^{|A|}\left(f((S \setminus B) \cup A_{i - 1} + a_i) - f((S \setminus B) \cup A_{i - 1}) - \alpha\right) \\ = f((S \setminus B) \cup A) - f(S \setminus B) - |A|\alpha \ge f(S \cup A) - f(S) - |A|\alpha
\end{multline}

\begin{algorithm}[t]
\caption{Non-Oblivious Local Search}
\label{alg:1}
\KwIn{\parbox[t]{7in}{\begin{itemize}[parsep=0em,itemsep=0em,topsep=0em,leftmargin=1em]
\item Ground set $\ground$ 
\item Membership oracle for $\mathcal{I} \subseteq 2^\ground$ 
\item Value oracle for monotone submodular function $f : 2^\ground \to \reals_+$ \item Approximation parameter $\epsilon \in (0,1)$
\end{itemize}}}
\DontPrintSemicolon
Let $\sinit = \{ \arg\max_{e \in \ground} f(\{e\} )\}$\;
Let $\delta = \left(1 + \frac{k + 3}{2\epsilon}\right)^{-1}$ and $\alpha = f(\sinit)\delta/n$\;
$S \gets \sinit$, $\prec\ \gets $ an arbitrary total ordering on $\ground$, and $\old{S} \gets S$\;
\Repeat{$\old{S} = S$}{
  Sort $S$ according to $\prec$ and let $s_i$ be the $i$th element in $S$\;
  $X \gets \emptyset$\;
  \For{$i = 1$ \KwTo $|S|$}
  {
    $w(s_i) \gets \displaystyle\left\lfloor (f(X + s_i) - f(X))/\alpha\right\rfloor\alpha$\;
    $X \gets X + s_i$\;
  }
  \ForEach{$k$ replacement $(A,B)$}{
    Sort $A$ according to $\prec$ and let $a_i$ be the $i$th element in $A$\;
    $X \gets S \setminus B$\;
    \For{$i = 1$ \KwTo $|A|$}{
      $w_{(A,B)}(a_i) \gets \displaystyle\left\lfloor (f(X + a_i) - f(X))/\alpha\right\rfloor\alpha$\;
      $X \gets X + a_i$\;
    }
    \If{$w_{(A,B)}^2(A) > w^2(S)$}{
      $\prec\ \gets $ the ordering $\prec'$ defined by
      $\begin{cases}
        x \prec' y & \text{for all $x \in S \setminus B$, $y \in A$}\\
        x \prec' y & \text{if $x \prec y$ for all other $x,y$}
      \end{cases}$\;
      $\old{S} \gets S$\;
      $S \gets (S \setminus B) \cup A$\;
      \Break\;
    }
  }
}
\Return $S$\;
\end{algorithm}

\section{Analysis of Algorithm \ref{alg:1}}

We now analyze the approximation and runtime performance of Algorithm \ref{alg:1}.  We now consider the worst-case ratio, or \emph{locality gap}, $f(O)/f(S)$ where $S$ is any locally optimal solution (with respect to Algorithm \ref{alg:1}'s potential function) and $O$ is a globally optimal solution.  We shall need the following technical lemma, which is a direct consequence of Lemma 1.1 in \cite{Lee-2010a}.  We give a proof here for the sake of completeness.

\begin{lemma}
\label{lem:submod}
Let $f$ be a submodular function on $\ground$, Let $T,S \subseteq \ground$, and $\{T_i\}_{i=1}^t$ be a partition of $T \setminus S$.  Then,
\[
\sum_{i=1}^t\left(f(S \cup T_i) - f(S)\right) \ge f(S \cup T) - f(S)
\]
\end{lemma}
\begin{proof}
Define $A_0 = S$ and $A_i = T_i \cup A_{i - 1}$ for all $1 \le i \le t$.  Suppose that $T_i = \{ t_j \}_{j = 1}^{|T_i|}$. Then, note that $S \subseteq A_{i - 1}$ and $T_i \cap A_{i - 1} = \emptyset$.  Submodularity of $f$ implies that:
\begin{align*}
f(T_i \cup S) - f(S) &= \sum_{j = 1}^{T_i}\left(f(\{t_l\}_{l = 1}^{j + 1} \cup S) - f(\{t_{l}\}_{l = 1}^j \cup S)\right) \\
&\ge  \sum_{j = 1}^{T_i}\left(f(\{t_l\}_{l = 1}^{j + 1} \cup A_{i - 1}) - f(\{t_{l}\}_{l = 1}^j \cup A_{i - 1})\right) \\
&= f(T_i \cup A_{i - 1}) - f(A_{i - 1}) = f(A_i) - f(A_{i - 1})
\end{align*}
Now, we have
\begin{equation*}
  \sum_{i = 1}^t\left(f(S \cup T_i) - f(S)\right) \ge \sum_{i = 1}^t\left(f(A_i) - f(A_{i - 1})\right) 
= f(A_t) - f(A_0) = f(S \cup T) - f(S) \qedhere
\end{equation*}
\end{proof}

We begin by considering the approximation ratio of Algorithm \ref{alg:1}.  Suppose that $S$ is the locally optimal solution returned by the algorithm on some instance, while $O$ is a global optimum for this instance.  Then, for every $k$-replacement $(A,B)$, we must have $w^2_{(A,B)}(A) \le w^2(B)$, where $w$ and each $w_{(A,B)}$ are weight functions determined by the solution $S$.  We consider only a particular subset of $k$-replacements in our analysis.

We have $S, O \in \I$ for the $k$-exchange system $\I$.   Thus, there must be a collection $Y$ assigning each $e$ of $O$ a neighborhood $Y_e \subseteq S$, satisfying the conditions of Definition \ref{def:k-exchange}.  For each $x \in S$, let $P_x$ be the set of all elements in $e \in O$ for which: (1) $x \in Y_e$ and (2) for all $z \in Y_e$, $w(z) \le w(x)$.  That is, $P_x$ is the set of all elements of $O$ in which $x$ is the heaviest element.  Note that the construction of $P_x$ depends on the fact that the weights $w$ assigned to elements in $S$ are \emph{fixed} throughout each iteration, and do \emph{not} depend on the particular improvement under consideration.  

We define $\N_x = \bigcup_{e \in P_x}Y_e$, and consider $(P_x, \N_x)$.  Property \ref{N2} of $Y$ ensures that $|P_x| \le k$.  Similarly, property \ref{N1}, together with the fact that all elements $e \in P_x$ have as a common neighbor $x \in Y_e$, ensures that $|\N_x| \le 1 + k(k - 1) = k^2 - k + 1$.  Finally, property \ref{N3} ensures that $(S \setminus \N_x) \cup P_x \in \I$.  Thus, $(P_x, \N_x)$ is a valid $k$-replacement for all sets $P_x \subseteq O$, $x \in S$.  Observe that $\{P_x\}_{x \in S}$ is a partition of $O$.  Furthermore, by the definition of $P_x$, we have $w(x) \ge w(z)$ for all $z \in \N_x$.  Again, this depends on the fact that the weights of elements in $S$ are the same for all $k$-replacements considered by the algorithm during a given phase.

The following extension of a theorem from \cite{Berman-2000} allows us to relate the non-oblivious potentials $w^2$ and $w^2_{(P_x,\N_x)}$ to the weight functions $w$ and $w_{(P_x,\N_x)}$ for each of our $k$-replacements $(P_x, \N_x)$.

\begin{lemma}
\label{thm:axy-lemma}
For all $x \in S$, and $e \in P_x$, 
$$w_{(P_x,\N_x)}^2(e) - w^2(Y_e - x) \ge w(x) \cdot \left(2 w_{(P_x,\N_x)}(e) - w(Y_e)\right) \enspace.$$
\end{lemma}
\begin{proof}
Let $a = \frac{1}{2}w(Y_e)$, and $b, c$ be such that
$w(x) = a + b$ and $w_{(P_x,\N_x)}(e) = a + c$ (note that $b$ and $c$ are not necessarily positive).  Then, since $e \in P_x$,
every element $z$ in $Y_e$ has weight at most $w(x) = a + b$.
Furthermore, $w(Y_e - x) = w(Y_e) - w(x) = a - b$.  Thus,
\begin{equation}
w^2(Y_e - x) = \sum_{z \in Y_e - x} w(z)^2 \le 
\sum_{z \in Y_e - x}(a + b) w(z) = (a + b)(a - b) \label{eq:6}
\end{equation}
Using \eqref{eq:6} and our definition of $a$, $b$, and $c$, we have
\begin{multline*}
 w_{(P_x,\N_x)}^2(e) - w^2(Y_e - x) - w(x) \cdot (2 w_{(P_x,\N_x)}(e) - w(Y_e)) \\
\ge (a + c)^2 - (a + b)(a - b) - (a + b)(2a + 2c - 2a)
= (b - c)^2 \ge 0 \enspace. \qedhere
\end{multline*}
\end{proof}

Using Lemma \ref{thm:axy-lemma} we can prove the following lemma, which uses the local optimality of $S$ to obtain a lower bound on the weight $w(x)$ of each element $x \in S$.

\begin{lemma}
For each $x \in S$, $w(x) \ge\sum\limits_{e \in P_x}
\left(2w_{(P_x,\N_x)}(e) - w(Y_e)\right)$.
\label{thm:charge}
\end{lemma}
\begin{proof}
Because $S$ is locally optimal with respect to $k$-replacements, including in particular $(P_x,\N_x)$, we must have:
\begin{equation}
  \label{eq:7}
w^2_{(P_x,\N_x)}(P_x) \le w^2(\N_x)
\end{equation}
First, we consider the case $w(x) = 0$.  Recall that all the weights produced by the algorithm are non-negative.  Because $w(x)$ is the largest weight in $\N_x$, we must have $w(e) = 0$ for all $e \in \N_x$ and so $w^2(\N_x) = 0$.  Moreover, (\ref{eq:7}) implies that $w^2_{(P_x,\N_x)}(P_x) = 0$ as well, and so, in particular, $w_{(P_x,\N_x)}(e) = 0$.  The claim then follows.

Now, suppose that $w(x) \neq 0$. From (\ref{eq:7}), together with the fact that $x \in Y_e$ for all $e \in P_x$, and the non-negativity of all the weights $w$, we have:
\begin{equation}
w_{(P_x,\N_x)}^2(P_x) \le w^2(\N_x) \le w^2(x) + \sum\limits_{e \in P_x} w^2(Y_e - x) \enspace.\label{eq:8}
\end{equation}
Rearranging \eqref{eq:8} using $w^2_{(P_x,\N_x)}(P_x) = \sum_{e \in P_x}
w^2_{(P_x,\N_x)}(e)$ we obtain:
\begin{equation}
\sum\limits_{e \in P_x} w^2_{(P_x,\N_x)}(e) - w^2(Y_e - x) \le w^2(x) \enspace.
\label{eq:9}
\end{equation}
Applying Lemma \ref{thm:axy-lemma} to 
each term on the
left of \eqref{eq:9}  we have:
\begin{equation}
\sum\limits_{e \in P_x} w(x)\cdot(2w_{(P_x,\N_x)}(e) - w(Y_e)) \le \sum\limits_{e \in P_x} w^2_{(P_x,\N_x)}(e) - w^2(Y_e - x)  \le w^2(x) = w(x)^2 \enspace.
\label{eq:19}
\end{equation}
Dividing by $w(x)$ (recall that $w(x) \neq 0$) then yields
\begin{equation*}
\sum\limits_{e \in P_x} \left(2w_{(P_x,\N_x)}(e) - w(Y_e)\right) \le w(x) \enspace.\qedhere
\end{equation*}
\end{proof}

We now prove our main result, which gives an upper bound on the locality gap of Algorithm \ref{alg:1}.

\begin{theorem}
\label{thm:locality-gap}
$\left( \frac{k + 3}{2} + \epsilon \right) f(S) \ge f(O)$
\end{theorem}
\begin{proof}
Lemma \ref{thm:charge} gives us one inequality for each $x \in S$. 
We now add all $|S|$ inequalities to obtain
\begin{equation}
  \sum_{x \in S}\sum_{e \in P_x}\left(2w_{(P_x,\N_x)}(e) -
  w(Y_e)\right) \le \sum_{x \in S} w(x) \enspace.
\label{eq:10}
\end{equation}
We have $\sum_{x \in S}w(x) \le f(S)$ by (\ref{eq:15r}).  Additionally, from (\ref{eq:16r}), $ f(S \cup P_x) - f(S) - |P_x|\alpha \le \sum_{e \in P_x} w_{(P_x,\N_x)}(e)$ for every $P_x$.  Thus, (\ref{eq:10}) implies
\begin{equation}
  2\sum_{x \in S}\left(f(S \cup P_x) - f(S) - |P_x|\alpha \right) - 
  \sum_{x \in S}\sum_{e \in P_x}w(Y_e) \le f(S) \enspace.
\label{eq:11}
\end{equation}
Since $\{P_x\}_{x \in S}$ is a partition of $O$, (\ref{eq:11}) is equivalent to
\begin{equation}
  2\sum_{x \in S}\left(f(S \cup P_x) - f(S)\right) - 2|O|\alpha 
  - \sum_{e \in O}w(Y_e) \le f(S) \enspace.
\label{eq:12}
\end{equation}
We  have $w(x) \ge 0$ for all $x \in S$, and there are at most $k$ distinct $e$ for which $x \in Y_e$, by property \ref{N2} of $Y$.  Thus, we have 
\[\sum_{e \in O}w(Y_e) \le k\sum_{x \in S}w(x) \le kf(S)\enspace,\] 
by (\ref{eq:15}).  Combining this with (\ref{eq:12}), we obtain
\begin{equation}
  2\sum_{x \in S}\left(f(S \cup P_x) - f(S)\right) - 2|O|\alpha - kf(S) \le f(S)
\label{eq:2}
\end{equation}
Using again the fact that $P$ is a partition of $O$, we can apply Lemma \ref{lem:submod} to the remaining sum on the left of \ref{eq:2}, yielding
\begin{equation*}
  2\left(f(S \cup O) - f(S)\right) - 2|O|\alpha - kf(S) \le f(S)\enspace
\end{equation*}
which simplifies to
\begin{equation}
\label{eq:17}
  f(S \cup O) - |O|\alpha \le \frac{k + 3}{2}f(S)\enspace.
\end{equation}
From the definition of $\alpha$ and the optimality of $O$, we have 
\begin{equation*}
\label{eq:1}
|O|\alpha \le n\alpha = \delta f(\sinit) \le \delta f(O)\enspace.
\end{equation*}
Finally, since $f$ is monotone, we have $f(S \cup O) \ge f(O)$.  Thus, (\ref{eq:17}) implies:
\begin{equation*}
(1 - \delta)f(O) \le \frac{k + 3}{2}f(S)\enspace,
\end{equation*}
which, after expanding the definition of $\delta$ and simplifying, is equivalent to $f(O) \le \left(\frac{k + 3}{2} + \epsilon\right)f(S)$.
\end{proof}

Next, we consider the runtime of Algorithm \ref{alg:1}.  Each iteration requires time 
$O(n)$ to compute the weights for $S$, plus time to evaluate all potential $k$-replacements.  There are $O(n^{k + k(k - 1) + 1}) = O(n^{k^2 + 1})$ such $k$-replacements $(A,B)$, and each one can be evaluated in time $O(k^2)$, including the computation of the weights $w_{(A,B)}$.  Thus, the total runtime of Algorithm \ref{alg:1} is $O(Ik^2n^{k^2 + 1})$, where $I$ is the number of improvements it makes.
The main difficulty remaining in our analysis is showing that Algorithm \ref{alg:1} constantly improves some global quantity, and so $I$ is bounded.  Here, we show that although the weights $w$ assigned to elements of $S$ change at each iteration, the non-oblivious potential $w^2(S)$, is monotonically increasing.  While the preceding analysis of the locality gap is valid regardless of the particular ordering $\prec$ used to generate the weights, our analysis of the convergence of Algorithm \ref{alg:1} requires that $\prec$ be updated at each phase to maintain the relative ordering of all elements in the current solution.

Finally, we consider what happens to the total squared weight $w^2(S)$ of the current solution after applying an $k$-replacement $(A,B)$.  In order to show that our algorithm terminates, we would like to show that this value is strictly increasing.  To show this, it is sufficient to show that each weight $w(x)$ for $x \in (S \setminus B) \cup A$ is strictly greater after applying the $k$-replacement than the corresponding weight before.  Unfortunately, the weight assigned to an element is highly sensitive to the ordering $\prec$ in which elements are considered.  Let $w$ be the weight function for solution $S$, and $w'$ be the weight function for solution $(S \setminus B) \cup A$.  If $x \prec y$ for some $x \in A$, $y \in S\setminus B$, then in the updated weight function, we could have $w'(y) < w(y)$, since $w'$ considers the marginal gain of $y$ with respect to a set containing $x$, while $w$ does not (since $x \not\in S$).  We avoid this phenomenon by updating the ordering $\prec$ each time an improvement is made.  In particular, we ensure that all the elements of $S$ and $A$ are considered in the same relative order, but that all of $A$ comes after all of $S$.  As we shall show, this ensures that the weights assigned each individual element in the $S \setminus B$ and $A$ solution do not decrease after applying the $k$-replacement $(A,B)$.

\begin{lemma}
Suppose that for some $k$-replacement $(A,B)$ and $\alpha > 0$, we have $w^2_{(A,B)}(A) \ge w^2(B) + \alpha$, and Algorithm \ref{alg:1} applies the replacement $(A,B)$ to $S$ to obtain solution $T = (S \setminus B) \cup A$.  Let $w_S$ be the weight function for solution $S$ and $w_T$ be the weight function for solution $T$.  Then, $w_T^2(T) \ge w^2_S(S) + \alpha$.
\label{lem:monotonic}
\end{lemma}
\begin{proof}
After applying the $k$-replacement $(A,B)$ to $S$, we obtain a new current solution $T = (S \setminus B) \cup A$ and a new ordering $\prec_T$.  We now show that for any element $x \in S \setminus B$, we must have $w(x) = w_S(x) \le w_T(x)$ and for any element $y \in A$, we must have $w_{(A,B)}(y) \le w_T(y)$.  

In the first case, let $S_x$ be the set of all elements in $S$ that come before $x$ in the ordering $\prec$, and similarly let $T_x$ be the set of all elements in $T$ that come before $x$ in $\prec_T$.  Suppose that for some element $z \in T$ we have $z \prec_T x$.  Then, since $x \in S \setminus B$, we must have $z \prec x$.  Thus, $T_x \subseteq S_x$.  It follows directly from the submodularity of $f$ that 
\begin{equation*}
w(x) = w_S(x) = \left\lfloor\frac{f(S_x + x) - f(S_x)}{\alpha}\right\rfloor\alpha \le \left\lfloor\frac{f(T_x + x) - f(T_x)}{\alpha}\right\rfloor\alpha = w_T(x)\enspace.
\label{eq:13}
\end{equation*}

In the second case, let $A_y$ be the set of all elements of $A$ that come before $y$ in the ordering $\prec$, and let $T_y$ be the set of all elements of $T$ that come before $y$ in the ordering $\prec_T$.  Suppose that for some element $z \in T$ we have $z \prec_T y$.  Then, since $y \in A$, we must have either $z \in S \setminus B$ or $z \in A$ and $z \prec y$.  Thus,
$T_y \subseteq (S \setminus B) \cup A_y$, and so
\begin{equation*}
w_{(A,B)}(y)\! =\! \left\lfloor\frac{f((S\! \setminus\! B) \cup A_y + y) - f((S\!\setminus\! B) \cup A_y)}{\alpha}\right\rfloor\!\alpha \le \left\lfloor\frac{f(T_y + y) - f(T_y)}{\alpha}\right\rfloor\!\alpha  = w_T(y) \enspace .
\label{eq:14}
\end{equation*}

From the above bounds on $w$ and $w_{(A,B)}$, together with the assumption of the lemma, we now have
\begin{multline*}
w^2_S(S) = \sum_{x \in S \setminus B}w_S(x)^2 + \sum_{x \in B}w_S(x)^2 
\le \sum_{x \in S \setminus B}w_S(x)^2 + \sum_{y \in A}w_{(A,B)}(y)^2 + \alpha
\\\le \sum_{x \in S \setminus B}w_T(x)^2 + \sum_{y \in A}w_{T}(y)^2 + \alpha
= w_T^2(T) + \alpha \enspace .\qedhere
\end{multline*}
\end{proof}

\begin{theorem}
\label{thm:runtime}
For any value $\epsilon \in (0,1)$, Algorithm \ref{alg:1} makes at most $O(n^3\epsilon^{-2})$ improvements.
\end{theorem}
\begin{proof}
Note submodularity implies that for any element $e$ and any set $T \subseteq \ground$, we must have $f(T + e) - f(T) \le f(\{e\}) \le f(\sinit)$.  In particular, for any solution $S \subseteq \ground$ with associated weight function $w$, we have
\[w^2(S) = \sum_{e \in S}w(e)^2 \le |S|f(\sinit)^2 \le nf(\sinit)^2\enspace.\]

Consider a given improvement $(A,B)$ applied by the algorithm.  Because every weight used in the algorithm is a multiple of $\alpha$, we have $w_{(A,B)}^2(A) > w^2(B)$ only if $w_{(A,B)}^2(A) \ge w^2(B) + \alpha^2$.  Let $T  = (S \setminus B) \cup A$ be the solution resulting from the improvement, and, as in the proof of Lemma \ref{lem:monotonic}, let $w_S$ be the weight function associated with $S$ and $w_T$ be the weight function associated with $T$.  For any $\epsilon > 0$, we have $\alpha > 0$, and hence $\alpha^2 > 0$.  Thus, from Lemma \ref{lem:monotonic}, after applying the improvement must we must have $w^2_T(T) \ge w^2_S(S) + \alpha^2$.

Thus, the number of improvements we can make is at most
\[\frac{nf(\sinit)^2 - f(\sinit)^2}{\alpha^2} = 
(n - 1)\left(\frac{f(\sinit)}{\alpha}\right)^2 = (n - 1)\frac{n^2}{\delta^2} = O(n^3\epsilon^{-2}) \enspace.\qedhere \]
\end{proof}

\begin{corollary}
For any $\epsilon > 0$, Algorithm \ref{alg:1} is a $\frac{k + 3}{2} + \epsilon$ approximation algorithm, running in time $O(\epsilon^{-2}k^2n^{k^2 + 4})$.
 \end{corollary}

\section{Open Questions}
\label{sec:open-questions}
We do not currently have an example for which the locality gap of Algorithm \ref{alg:1} can be as bad as stated, even for specific $k$-exchange systems such as $k$-set packing.  
In the particular case of weighted independent set in $(k + 1)$-claw free graphs Berman \cite{Berman-2000} gives a tight example that shows his algorithm can return a set $S$ with $\frac{k + 1}{2}w(S) = w(O)$.  His example uses only unit weights, and so the non-oblivious potential function is identical to the oblivious one.  However, the algorithm of Feldman et al.\ (given here as Algorithm \ref{alg:nols}) considers a larger class of improvements than those considered by  Berman, and so Berman's tight example no longer applies, even in the linear case.  
For the unweighted variant, Hurkens and Schrijver give a lower bound of $k/2 + \epsilon$, where $\epsilon$ depends on the size of the improvements considered.  Because the non-oblivious local search routine performs the same as oblivious local search on instances with unit weights (since $1 = 1^2$), this lower bound applies to Algorithm \ref{alg:nols} in the linear case.  
From a hardness perspective, the best known bound is the $\Omega(k/\ln k)$ NP-hardness result of Hazan, Safra, and Schwartz \cite{Hazan-2006}, for the special case of unweighted $k$-set packing.  

In addition to providing a tight example for our analysis, it would be interesting to see if similar techniques could be adapted to apply to more general problems such as matroid $k$ parity in arbitrary matroids (here, even an improvement over $k$ for the general linear case would be interesting) or to non-monotone submodular functions.  A major difficulty with the latter generalization is our proof's dependence on the weights' non-negativity, as this assumption no longer holds if our approach is applied directly to non-monotone submodular functions.

\subparagraph*{Acknowledgment}

The author thanks Allan Borodin for providing comments on a preliminary version of this paper.

\bibliographystyle{abbrv}
\bibliography{PapersNew}

\begin{thebibliography}{10}

\bibitem{Arkin-1997}
E.~M. Arkin and R.~Hassin.
\newblock On local search for weighted k-set packing.
\newblock In {\em Proc. of 5th ESA}, pages 13--22, 1997.

\bibitem{Berman-2000}
P.~Berman.
\newblock A d/2 approximation for maximum weight independent set in d-claw free
  graphs.
\newblock {\em Nordic J. of Computing}, 7:178--184, Sept. 2000.

\bibitem{Brualdi-1971}
R.~A. Brualdi.
\newblock Induced matroids.
\newblock {\em Proc. of the American Math. Soc.}, 29:213--221, 1971.

\bibitem{Calinescu-2007}
G.~Calinescu, C.~Chekuri, M.~P{\'a}l, and J.~Vondr{\'a}k.
\newblock Maximizing a submodular set function subject to a matroid constraint.
\newblock In {\em Proc. of 12th IPCO}, pages 182--196, 2007.

\bibitem{Chandra-1999}
B.~Chandra and M.~Halld\'{o}rsson.
\newblock Greedy local improvement and weighted set packing approximation.
\newblock In {\em Proc. of 10th ACM-SIAM SODA}, pages 169--176, 1999.

\bibitem{Feige-1998}
U.~Feige.
\newblock A threshold of ln n for approximating set cover.
\newblock {\em J. {ACM}}, 45:634--652, July 1998.

\bibitem{Feige-2007}
U.~Feige, V.~S. Mirrokni, and J.~Vondrak.
\newblock Maximizing non-monotone submodular functions.
\newblock In {\em Proc. of 48th IEEE FOCS}, pages 461--471, 2007.

\bibitem{Feldman-2011a}
M.~Feldman, J.~S. Naor, and R.~Schwartz.
\newblock Nonmonotone submodular maximization via a structural continuous
  greedy algorithm.
\newblock In {\em Proc. of 38th ICALP}, pages 342--353, 2011.

\bibitem{Feldman-2011}
M.~Feldman, J.~S. Naor, R.~Schwartz, and J.~Ward.
\newblock Improved approximations for $k$-exchange systems.
\newblock In {\em Proc. of 19th ESA}, pages 784--798, 2011.

\bibitem{Fisher-1978}
M.~L. Fisher, G.~L. Nemhauser, and L.~A. Wolsey.
\newblock An analysis of approximations for maximizing submodular set
  functions---{II}.
\newblock In {\em Polyhedral Combinatorics}, volume~8 of {\em Mathematical
  Programming Studies}, pages 73--87. Springer Berlin Heidelberg, 1978.

\bibitem{Gharan-2011}
S.~O. Gharan and J.~Vondr\'{a}k.
\newblock Submodular maximization by simulated annealing.
\newblock In {\em Proc. of 22nd ACM-SIAM SODA}, pages 1098--1116, 2011.

\bibitem{Gupta-2010}
A.~Gupta, A.~Roth, G.~Schoenebeck, and K.~Talwar.
\newblock Constrained non-monotone submodular maximization: Offline and
  secretary algorithms.
\newblock In {\em Proc. of 7th WINE}, pages 246--257, 2010.

\bibitem{Halldorsson-1995}
M.~M. Halld\'{o}rsson.
\newblock Approximating discrete collections via local improvements.
\newblock In {\em Proc. of 6th ACM-SIAM SODA}, pages 160--169, 1995.

\bibitem{Hazan-2006}
E.~Hazan, S.~Safra, and O.~Schwartz.
\newblock On the complexity of approximating k-set packing.
\newblock {\em Computational Complexity}, 15:20--39, May 2006.

\bibitem{Hurkens-1989}
C.~A.~J. Hurkens and A.~Schrijver.
\newblock On the size of systems of sets every t of which have an sdr, with an
  application to the worst-case ratio of heuristics for packing problems.
\newblock {\em SIAM J. Discret. Math.}, 2(1):68--72, 1989.

\bibitem{Khanna-1994}
S.~Khanna, R.~Motwani, M.~Sudan, and U.~Vazirani.
\newblock On syntactic versus computational views of approximability.
\newblock In {\em Proc. of 35th IEEE FOCS}, pages 819--830, 1994.

\bibitem{Lee-2009}
J.~Lee, V.~S. Mirrokni, V.~Nagarajan, and M.~Sviridenko.
\newblock Non-monotone submodular maximization under matroid and knapsack
  constraints.
\newblock In {\em Proc. 41st ACM STOC}, pages 323--332, 2009.

\bibitem{Lee-2010}
J.~Lee, M.~Sviridenko, and J.~Vondr\'{a}k.
\newblock Matroid matching: the power of local search.
\newblock In {\em Proc. of 42nd {ACM} STOC}, pages 369--378, 2010.

\bibitem{Lee-2010a}
J.~Lee, M.~Sviridenko, and J.~Vondr{\'a}k.
\newblock Submodular maximization over multiple matroids via generalized
  exchange properties.
\newblock {\em Math. of Oper. Res.}, 35(4):795 --806, Nov. 2010.

\bibitem{Soto-2011a}
J.~A. Soto.
\newblock A simple {PTAS} for weighted matroid matching on strongly base
  orderable matroids.
\newblock {\em Electronic Notes in Discrete Mathematics}, 37:75--80, Aug. 2011.

\end{thebibliography}
\end{document}